\documentclass[pre,aps,superscriptaddress,nofootinbib]{revtex4-2}
\usepackage{amsfonts}
\usepackage{times}
\usepackage{amsmath}
\usepackage{graphicx}
\usepackage{amssymb}
\usepackage{xcolor,comment}

\usepackage{caption}
\usepackage{subcaption}
\usepackage{algorithm}
\usepackage{algpseudocode}

\begin{document}

\title{Discrete Breathers in Klein--Gordon Lattices: a Deflation-Based Approach}

\author{F.~Martin-Vergara}
\affiliation{\'{A}rea B\'{a}sica de Tecnolog\'{i}as de la Informaci\'{o}n y Comunicaciones, Servicio de Sistemas Inform\'{a}ticos.
Universidad de Málaga, 29071-M\'alaga, Spain}
\author{J.~Cuevas-Maraver}
\affiliation{Grupo de F\'{\i}sica No Lineal, Departamento de F\'{\i}sica Aplicada I,
Universidad de Sevilla. Escuela Polit\'{e}cnica Superior, C/ Virgen de \'{A}frica, 7, 41011-Sevilla, Spain}
\affiliation{Instituto de Matem\'{a}ticas de la Universidad de Sevilla (IMUS). Edificio
Celestino Mutis. Avda. Reina Mercedes s/n, 41012-Sevilla, Spain}
\email{jcuevas@us.es}
\author{P.~E.~Farrell}
\email{patrick.farrell@maths.ox.ac.uk}
\affiliation{Mathematical Institute, University of Oxford, Oxford OX2 6GG, United Kingdom}
\author{F.~R.~Villatoro}
\affiliation{Escuela de Ingenier\'{\i}as Industriales, Departamento de Lenguajes y Ciencias de la Computaci\'on,
Universidad de Málaga, 29071-M\'alaga, Spain}
\email{frvillatoro@uma.es}
\author{P.~G.~Kevrekidis}
\affiliation{Department of Mathematics and Statistics, University
of Massachusetts, Amherst, Massachusetts 01003-4515, USA}

\begin{abstract}
Deflation is an efficient numerical technique for identifying new branches of steady state solutions to nonlinear
partial differential equations. Here, we demonstrate
how to extend deflation to discover new
periodic orbits in nonlinear dynamical lattices. We employ our extension to identify
discrete breathers,
which are generic exponentially localized, time-periodic
solutions of such lattices. We compare
different approaches to using deflation for periodic orbits,
including ones based on a Fourier decomposition
of the solution, as well as ones based on the
solution's energy density profile. We demonstrate
the ability of the method to obtain a wide
variety of multibreather solutions without prior
knowledge about
their spatial profile.
\end{abstract}

\maketitle

\section{Introduction}

The study of discrete breathers has received extensive attention over
the past four decades and has been
summarized in numerous
reviews~\cite{willis,Flach2008,Aubry2006}.
Part of their appeal can be attributed to their
broad relevance: they generically appear in
anharmonic crystals, as argued
since the early works of~\cite{takeno,page}.
Their mathematical existence is rigorously established
(under mild non-resonance
conditions) in nonlinear dynamical
lattices~\cite{mackay}.
These features were, in turn,
mirrored in their demonstrated experimental emergence
in a wide range of applications, such as the nonlinear
dynamics of optical waveguides~\cite{LEDERER20081},
the setting of mean-field atomic condensates
in optical lattices~\cite{RevModPhys.78.179},
the evolution of granular crystals in materials
science~\cite{Chong2018},
electrical circuits~\cite{remoissenet},
Josephson-junction ladders~\cite{alex,alex2}, micromechanical arrays \cite{cantilevers},
and models of the
DNA double strand \cite{Peybi}, crystals \cite{NbNi} or carbon materials \cite{carbon}.

In this work we will extend the idea of deflation~\cite{farrell2015deflation}
to discovering discrete breathers.
Deflation is a numerical technique for computing multiple solutions of nonlinear equations.
Once a solution of the equations has been found (e.g.~with Newton's method), the residual of the problem is modified so that the known solution
is removed: Newton's method will not converge to it again, under mild regularity conditions. Newton's method may then be applied again,
and if it converges it will have discovered a second solution. The process may then be repeated to yield multiple solutions from the same initial guess.
The technique demonstrates exciting potential
for unlocking a wide
range of unknown solutions, importantly
without expert knowledge of the system statics
or dynamics. Recently, the technique has
been shown to enable the identification of
a wide range of steady state solutions
in dispersive nonlinear systems, both in two and three dimensions and for both
single and multiple
components~\cite{charalampidis2018computing,charalampidis2019bifurcation,boulle2020deflation,stereo}.

The use of deflation has, for
the most part, been limited to the identification
of stationary states, while attempts to apply
it to the realm of periodic orbits have
been very limited; see, e.g.,~\cite{vraha} for
a low-dimensional example. Indeed, we show
that such a technique can be brought to bear
towards the identification of periodic orbits
in the form of discrete breathers for high-dimensional
nonlinear lattice dynamical systems.

 More specifically, we illustrate
 that, upon taking
the Fourier series decomposition of the periodic
solution, the problem acquires an algebraic
form (effectively in ``space-time''), where
deflation can be directly applied. Indeed, we
offer multiple variants of this, one in which
the vector of associated Fourier coefficients
is deflated against, and another where
the energy density profile is used instead.
The latter is commonly (although not necessarily)
stationary during the periodic (amplitude) variation of
discrete breathers. We show that the technique
is successful in uncovering a wide range
of previously unknown solutions (to the best
of our knowledge) in a Klein--Gordon lattice bearing
an onsite nonlinear potential of the Morse type.
While many of these solutions could have been
constructed by leveraging expert knowledge, e.g.,
from the uncoupled so-called anti-continuum limit~\cite{mackay}, the present method allows
us to automatically obtain them, without such prior
knowledge. Hence, deflation and the
associated deflated continuation are argued
to be an efficient tool for unraveling the landscape
of periodic orbits in (many-degree-of-freedom)
lattice nonlinear dynamical systems.

Our presentation is structured as follows.
In section II we provide details of the
theory and computational method
associated with the use of deflation for
periodic orbits. In section III, we
explore the Klein--Gordon lattice associated
with the onsite Morse potential. We apply deflation
in combination with continuation to systematically
obtain the periodic orbits of the system.
Finally, in section IV, we summarize our
findings and propose a number of directions
for future study.

\section{Theoretical and Computational Setup}

Our system of choice will be a Klein--Gordon chain of oscillators of the form:
\begin{equation}
    \label{eq:KG}
    \ddot{u}_n+V'(u_n)-C(u_{n-1}+u_{n+1}-2u_n)=0\,,\qquad n=\lfloor-(N-1)/2\rfloor\ldots \lfloor(N-1)/2\rfloor,
\end{equation}
with $u_n$ being the displacement from the equilibrium of the $n$-th oscillator, the double dot meaning the second derivative in time, $V(u_n)$ being the on-site (or substrate) potential and $C$ the coupling constant~\cite{Flach2008,Aubry2006}. This equation can be derived from the Hamiltonian:
\begin{equation}
    \label{eq:ham}
    H(\{u_n(t)\})=\sum_n \frac{\dot{u}^2_n}{2}+V(u_n)+\frac{C}{2}(u_n-u_{n+1})^2.
\end{equation}
We will seek time-reversible periodic solutions of equation \eqref{eq:KG}. To this aim, we express $u_n(t)$ as a truncated Fourier series expansion:
\begin{equation}
    \label{eq:Fourier}
    u_n(t)=z_{0,n}+2\sum_{k=1}^{K}z_{k,n}\cos(k\omega t),
\end{equation}
so that the set of $N$ ODEs transforms into a set of $(K+1)N$ nonlinear algebraic equations $\mathbf{F}(\{z_{k,n}\})=0$ that can be solved by fixed point methods.
This treatment of frequencies and
space on an equal footing for the resulting Fourier
coefficients yields a conversion of the system into
an algebraic one, which enables the use of
deflation. The
resulting equations take the form
\begin{equation}
    \label{eq:KG_Fourier}
    F_{k,n} \equiv -k^2\omega^2z_{k,n}+{\cal F}_{k,n}-C(z_{k,n+1}+z_{k,n-1}-2z_{k,n})=0.
\end{equation}
Here, ${\cal F}_{k,n}$ denotes the $k$-th
mode at the $n$-th site of the discrete cosine Fourier transform:
\begin{equation}
    {\cal F}_{k,n}=\frac{1}{2K+1}\left[V'(u_n(0))+2\sum_{q=1}^K V'(u_n(t_q))\cos(k\omega t_q)\right],
\end{equation}
with $u_n(t)$ taken from \eqref{eq:Fourier} and $t_q=2\pi q/((2K+1)\omega)$.

One can acquire a discrete breather solution by making use of continuation from the anti-continuum limit introduced by MacKay and Aubry~\cite{mackay}. The
relevant theorem establishes that a solution at the $C=0$ limit can be continued up to a finite coupling as long as no integer multiple of the breather frequency $\omega$ resonates with the linear  band modes. With this in mind, it is possible to compute the solution for a single oscillator $\tilde{u}(t)$ and construct the breather (centered at $n=0$) at the anti-continuum limit as $u_n(t)=\tilde{u}(t)\delta_{n,0}$. This solution can then be continued, e.g., through the Newton--Raphson method until a prescribed coupling constant~\cite{Marin}; for a single-site excitation
such a continuation will generically exist.
For the multi-breather structures considered below
such a continuation is less straightforward, as will be
seen.
For an infinite oscillator chain, the continuation finishes when $n\omega$, with $n\in\mathbf{N}$, coincides with one of the borders of the linear modes band \cite{Aubry97}. If the chain is finite, there are gaps in the linear modes spectrum allowing the existence of breathers for which $n\omega$ is inside the linear modes band \cite{phantom}.

There are many ways of constructing breathers at the anti-continuum limit, depending on the sites where the isolated oscillator is located and the phase of those oscillators. Breathers with more than one excited site are usually denoted as multibreathers~\cite{Aubry2006,Juan,Koukouloyannis_2009,Sakovich}. Among them, the only one that can be continued until the resonance with linear modes is the so-called inter-site mode (while the on-site mode, mentioned in the paragraph above, concerns a single excited site), which is given by $u_n(t)=\tilde{u}(t)(\delta_{n,0}+\delta_{n,1})$, i.e., it consists of two adjacent excited sites oscillating in phase\footnote{It is
well-known from works such as those of~\cite{Juan,Koukouloyannis_2009,Sakovich} that such
solutions are dynamically unstable in Klein--Gordon
lattices with soft nonlinearities such as the
ones considered herein.}. The other multibreathers experience a bifurcation when one performs continuation on $C$ before the resonance with linear modes.

We will consider breathers in the Klein--Gordon chain with the Morse potential~\cite{Peybi}, $V(u)=D(\mathrm{e}^{-bx}-1)$, with $D=1/2$ and $b=1$, and a fixed frequency $\omega=0.8$. The coefficients of the Fourier series for an isolated oscillator can be analytically calculated (see e.g. \cite[Appendix A]{Juan}), so that $\tilde{u}=\tilde{z}_0+2\sum_k \tilde{z}_k\cos(k\omega t)$ with
\begin{equation}
    \tilde{z}_0=\frac{\log(1+\omega)}{2\omega},\qquad \tilde{z}_k=-\frac{(-1)^k}{k}\left(\frac{1-\omega}{1+\omega}\right)^{k/2}.
\end{equation}
Upon regular (parametric) continuation, we extend solutions from $C=0$ to a given $C_{\textrm{max}}$ by using the solution at $C-\delta C$ as initial guess for Newton's method applied at $C$. In other words, we make small steps in parameter space, so that the desired solution is close to the initial guess. This allows us to compute a branch of on-site breathers for a relatively large value of $C_{\textrm{max}}$.

Deflation will subsequently
enable us to discover other branches of solutions, potentially quite far away in configuration space. The key idea is to solve a \emph{judiciously modified} equation $\mathbf{G} = 0$ that removes the known solutions from consideration.
Given the residual $\mathbf{F}$ for fixed parameters (as defined by \eqref{eq:KG_Fourier}), and
known solutions $z^{(\mu)}$, $\mu = 1, \dots, \nu$, we consider two variants for defining $\mathbf{G}$.
The first one
is based on the Fourier coefficients of the equations:
\begin{equation}
    \label{eq:deflation_FC}
    G_{k,n}\equiv \prod_{\mu=0}^{\nu}\left(\frac{1}{||z-z^{(\mu)}||_{l^p}}+\sigma\right)F_{k,n}
\end{equation}
Here $\sigma>0$ is a constant (typically $\sigma = 1$).  The prefactor multiplying the vector $\mathbf{F}$ is always strictly positive, so the only solutions to the vector equations
$\mathbf{G} = 0$ are those for which $\mathbf{F} = 0$. The prefactor blows up as $z$ approaches a known solution, which forces Newton-type methods not to converge there. Given $\nu \ge 0$ solutions, we can solve $\mathbf{G} = 0$ from any available initial guess (so long as it is not a known solution), and if this solve is successful we have discovered the $(\nu + 1)^\mathrm{th}$ solution.
We fix $z^{(0)}\equiv0$, to also deflate the trivial zero solution. In our work, we solve $\mathbf{G} = 0$ by means of a Levenberg--Marquardt algorithm and refined by the trust-region-dogleg algorithm
(as implemented in Matlab's \texttt{fsolve}).

Notice that this procedure may yield solutions that are related to known ones by group actions, e.g.~by time or spatial reversal or even translational shift.
In principle, this can be avoided by
building into deflation the symmetries of the original
dynamical system, as was done, e.g., in~\cite{charalampidis2019bifurcation}
in the continuum context.
This aspect is worth pursuing systematically in
the discrete realm in the future.

The above approach constitutes the ``standard''
deflation perspective for systems of algebraic
equations, relying on the use of the
Fourier decomposition to turn the computation
of the periodic orbit into an algebraic problem
in the two-dimensional space of frequencies
and spatial lattice nodes.
The second alternative for defining $\mathbf{G}$ is based on the breather energy density. In that case, we define
\begin{equation}
    \label{eq:deflation_energy}
    G_{k,n}\equiv \prod_{\mu=0}^{\nu}\left(\frac{1}{||E(z)-E(z^{(\mu)})||_{l^p}}+\sigma\right)F_{k,n}
\end{equation}
with $E(z)$ being given as the expression inside
the summation within
Eq.~\eqref{eq:ham},
i.e.,
\begin{equation}
    \label{eq:energy}
    E(z)=V(u_n(0))+\frac{C}{2}(u_n(0)-u_{n+1}(0))^2,\qquad u_n(0)=z_{n,0}+2\sum_{k=1}^K z_{n,k}
\end{equation}
Part of the motivation for considering this alternative stems from the
fact that often (but not always) the energy density of a breather, while
exponentially localized in space around the breather location, may be
independent of time. Hence it may also be a suitable vector entity to ``deflate
against''.

We now turn to the specifics of our numerical results to explore
the different types of multibreather solutions that one can obtain
starting from a fundamental solution of the nonlinear problem
(such as a single-site breather).

\section{Numerical Results}

Hereafter,
we will apply deflation at a particular value of the coupling, namely $C=0.05$. This choice is motivated for the great number of multibreathers existing for such a relatively low coupling.
This leads to a more complex energy landscape: as the coupling is
increased (as we will see also below) many of these solutions terminate
in different types of bifurcations, reducing the complexity of available
waveforms.

\begin{algorithm}[H]
  \begin{algorithmic}
  \For{case in Tables I \& II}
    \State known solutions $\gets [z^{(1)}, 0]$;
    \State perturb $z^{(1)}$ to get $z_\text{ptb}$;
    \State solutions found $\gets 0$;
    \While{true}
      \State solve deflated problem from initial guess $z_\text{ptb}$;
      \If{solve was successful}
        \State refine discovered solution by solving undeflated problem;
        \State append refined solution to known solutions;
        \State increment solutions found;
        \If{solutions found $= 10$}
          \State \Return known solutions.
        \EndIf
      \Else
        \State \Return known solutions.
      \EndIf
    \EndWhile
  \EndFor
  \end{algorithmic}
  \caption{Outline of the approach to finding multiple solutions.}
  \label{alg:approach}
\end{algorithm}

\begin{figure}[!ht]
\begin{center}
\begin{tabular}{cc}
\includegraphics[width=.45\textwidth]{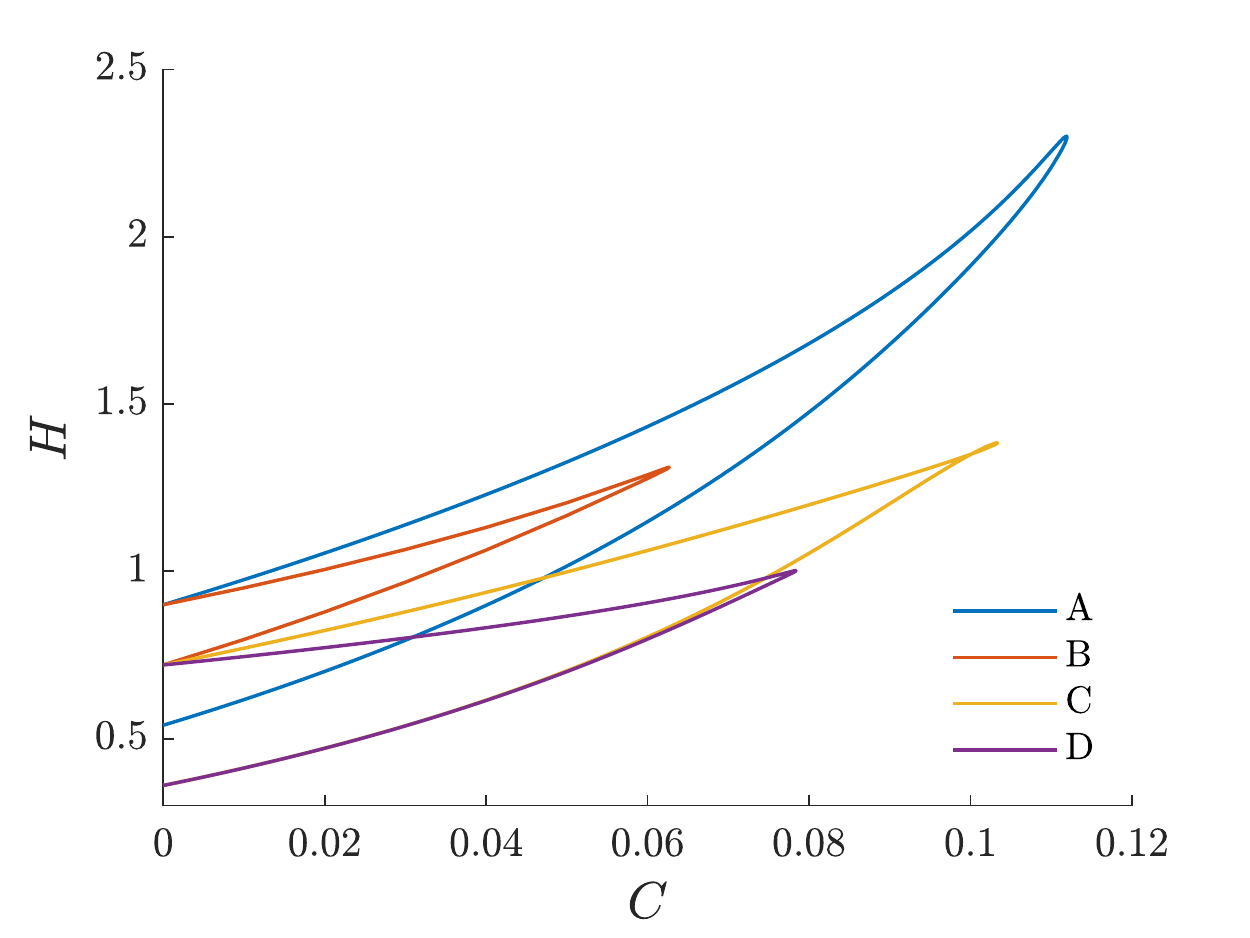} &
\includegraphics[width=.45\textwidth]{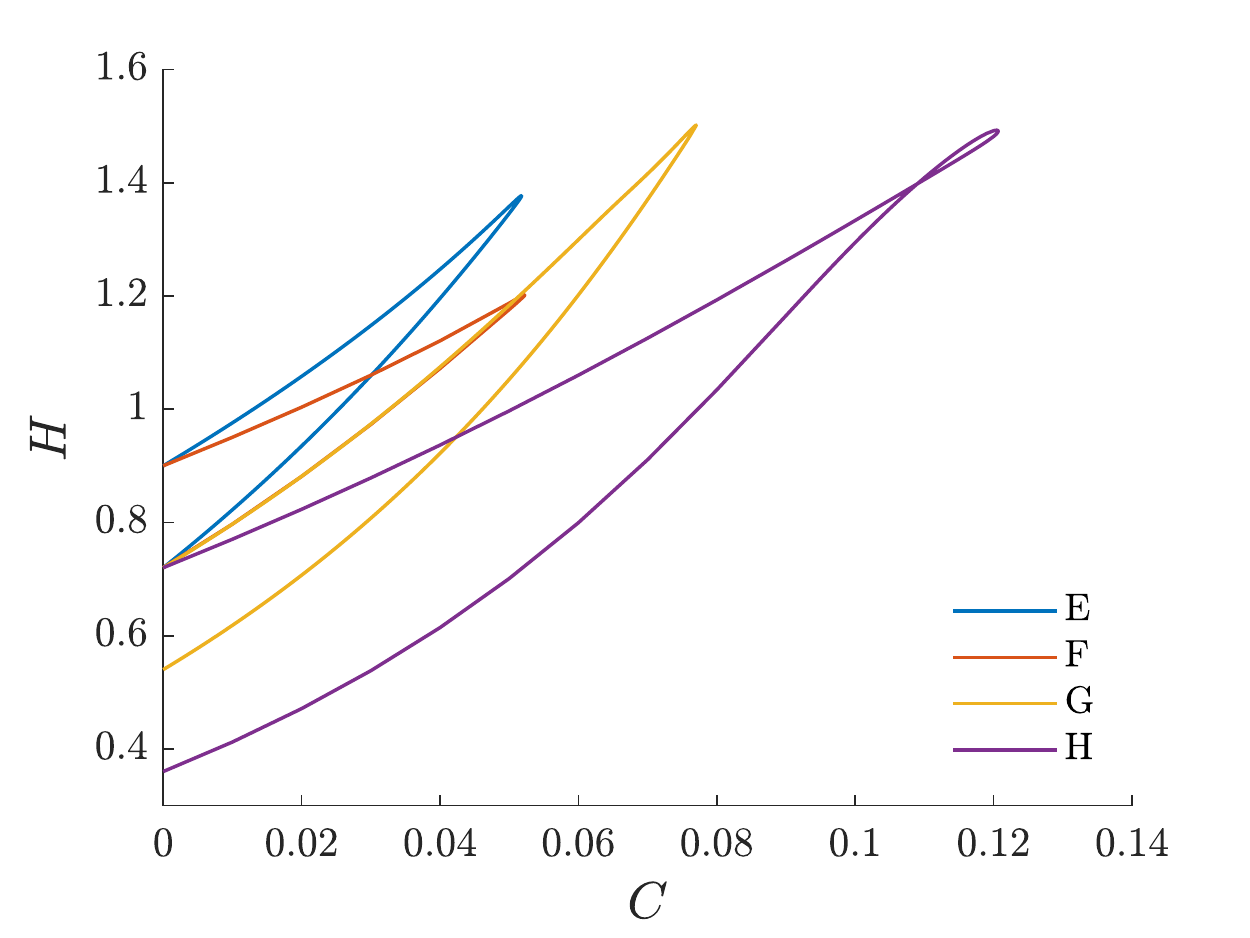} \\
\includegraphics[width=.45\textwidth]{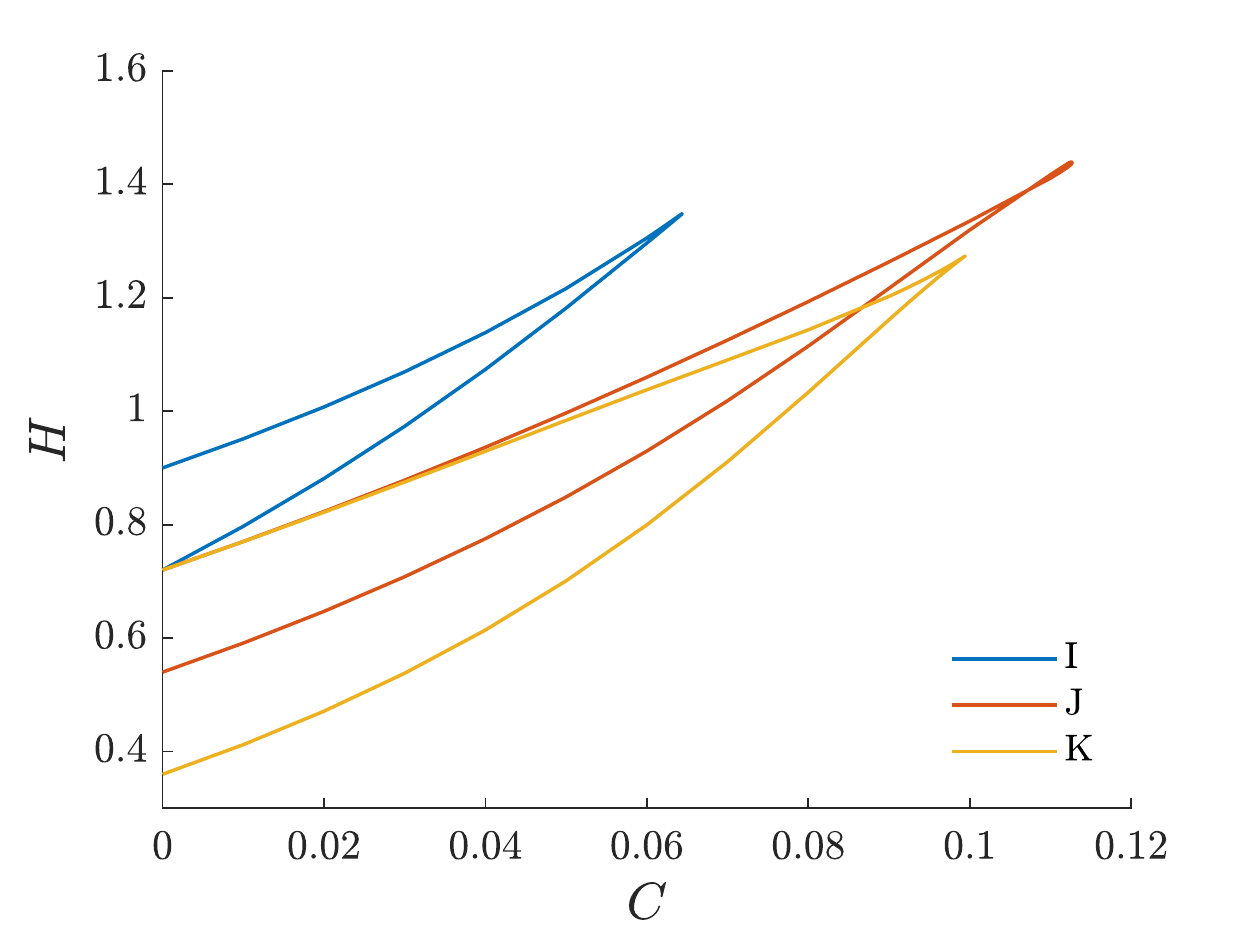} &
\includegraphics[width=.45\textwidth]{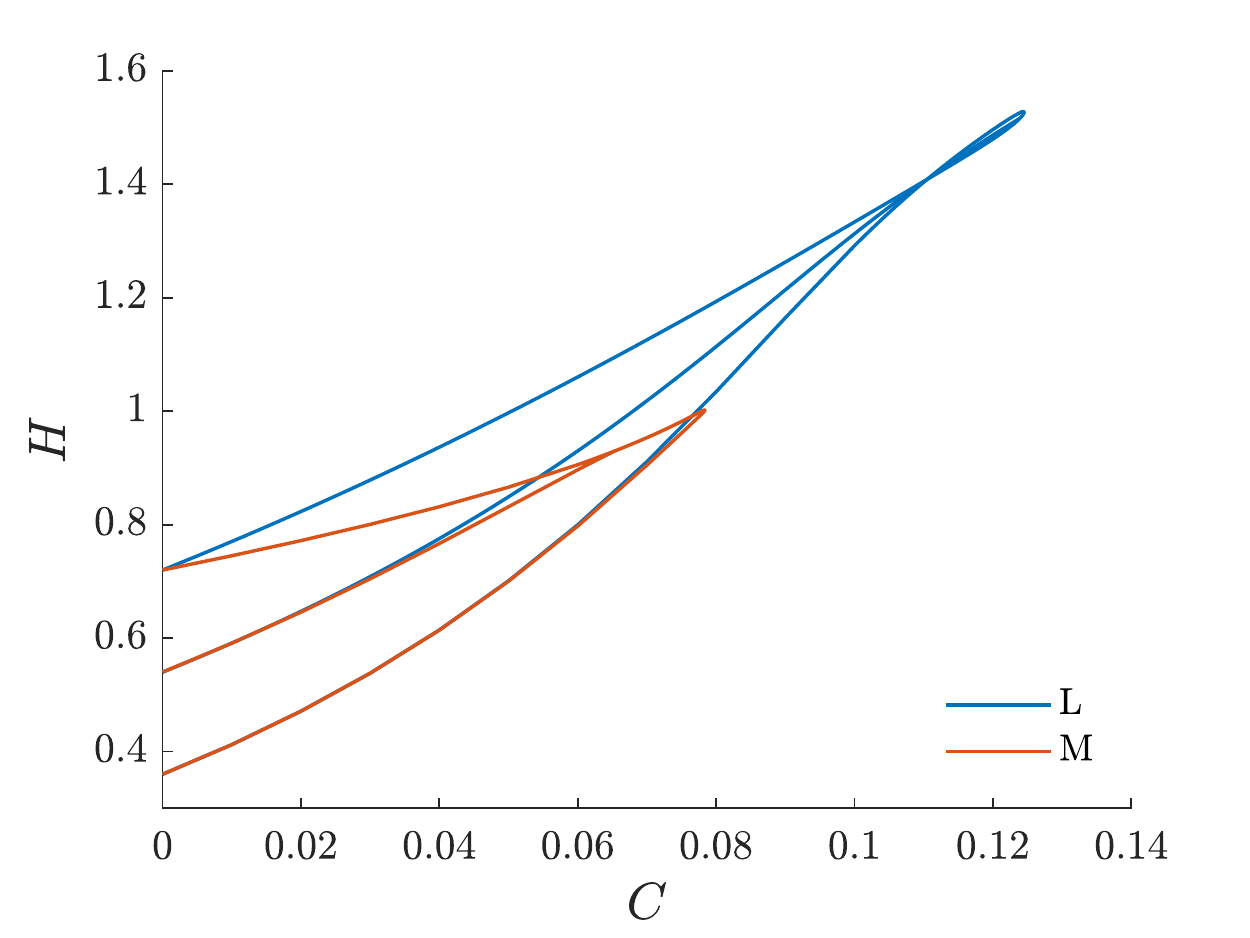} \\
\end{tabular}
\end{center}
\caption{Bifurcation diagrams for breather families obtained from a deflation perturbing the breather core. See the subsequent figures for elemental
members of each of the labeled solution families.}
\label{fig:diagrams1}
\end{figure}

\begin{figure}[!ht]
\begin{center}
\includegraphics[width=.9\textwidth]{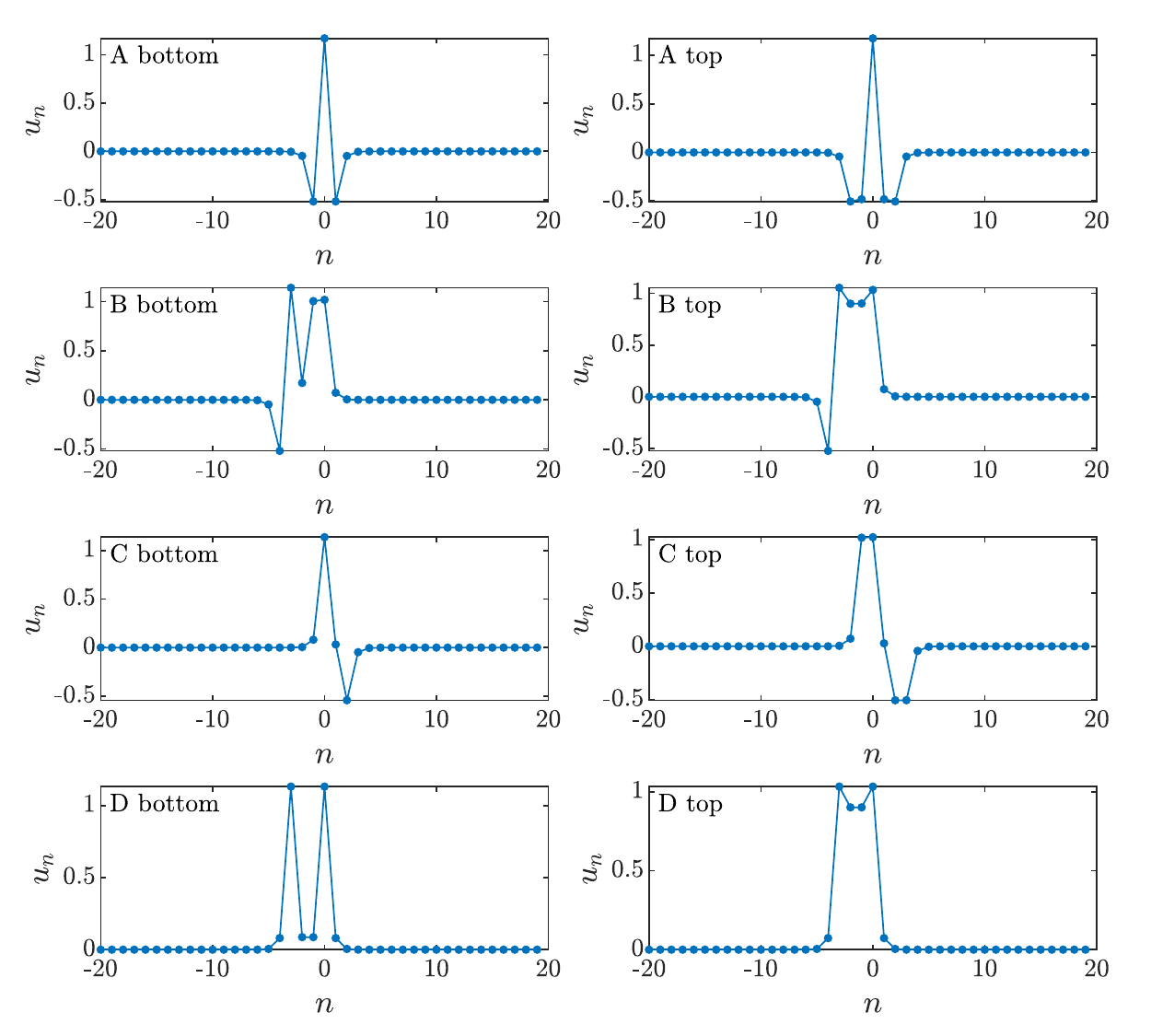}
\end{center}
\caption{Breather profiles for the solutions with $C=0.03$ in the upper left panel of Fig.~\ref{fig:diagrams1}.}
\label{fig:breathers1a}
\end{figure}

\begin{figure}[!ht]
\begin{center}
\includegraphics[width=.9\textwidth]{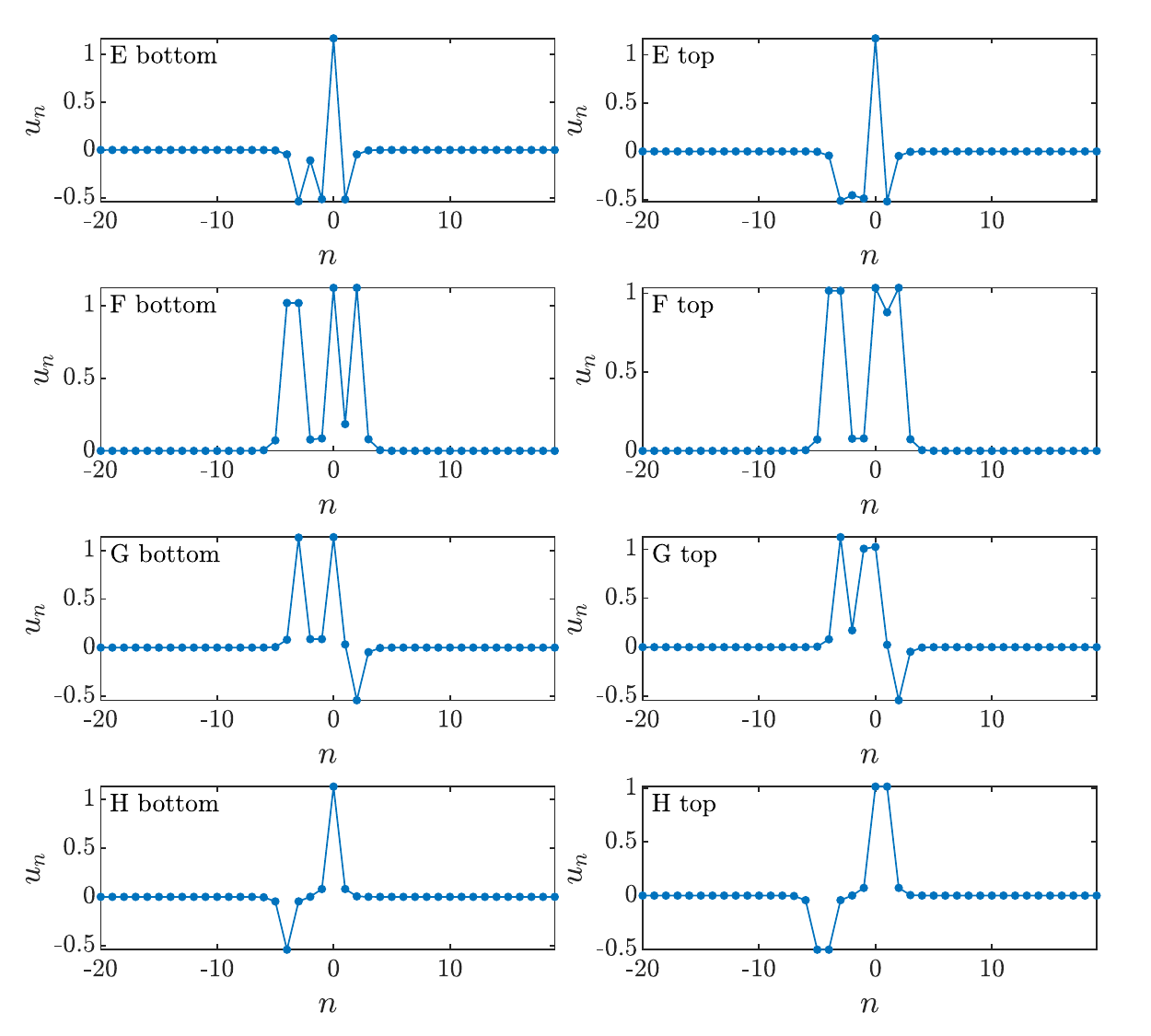}
\end{center}
\caption{Breather profiles for the solutions with $C=0.03$ in the upper right panel of Fig.~\ref{fig:diagrams1}.}
\label{fig:breathers1b}
\end{figure}

\begin{figure}[!ht]
\begin{center}
\includegraphics[width=.9\textwidth]{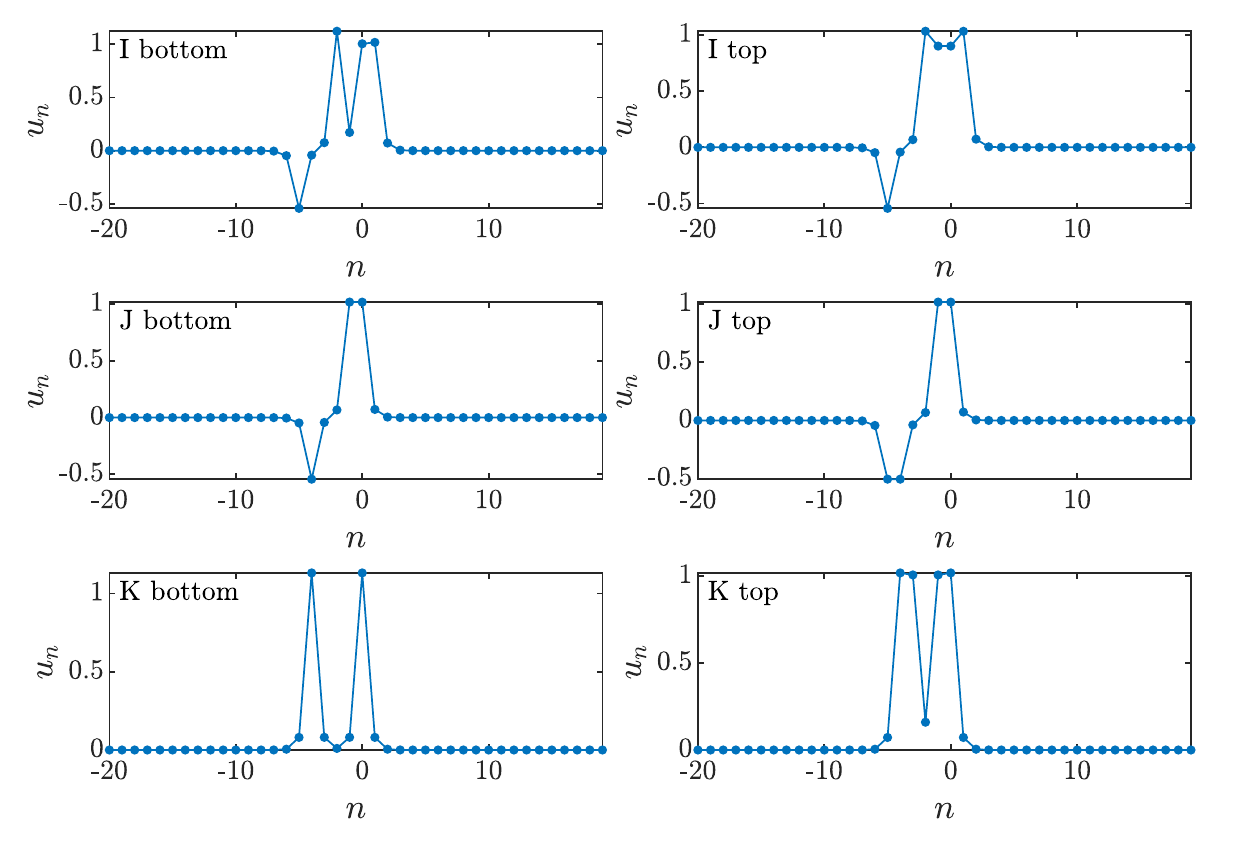}
\end{center}
\caption{Breather profiles for the solutions with $C=0.03$ in the bottom left panel of Fig.~\ref{fig:diagrams1}.}
\label{fig:breathers1c}
\end{figure}

\begin{figure}[!ht]
\begin{center}
\includegraphics[width=.9\textwidth]{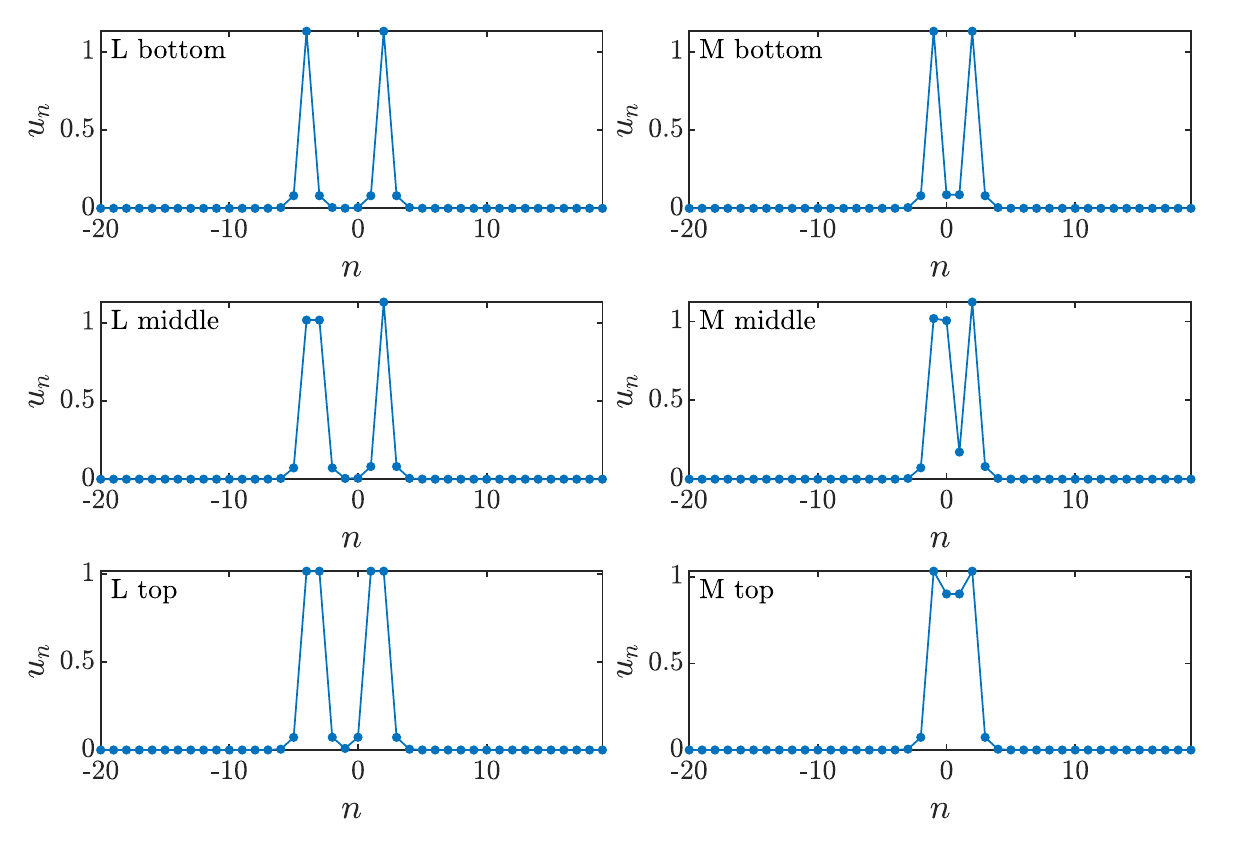}
\end{center}
\caption{Breather profiles for the solutions with $C=0.03$ in the bottom right panel of Fig.~\ref{fig:diagrams1}.}
\label{fig:breathers1d}
\end{figure}

\begin{figure}[!ht]
\begin{center}
\begin{tabular}{cc}
\includegraphics[width=.45\textwidth]{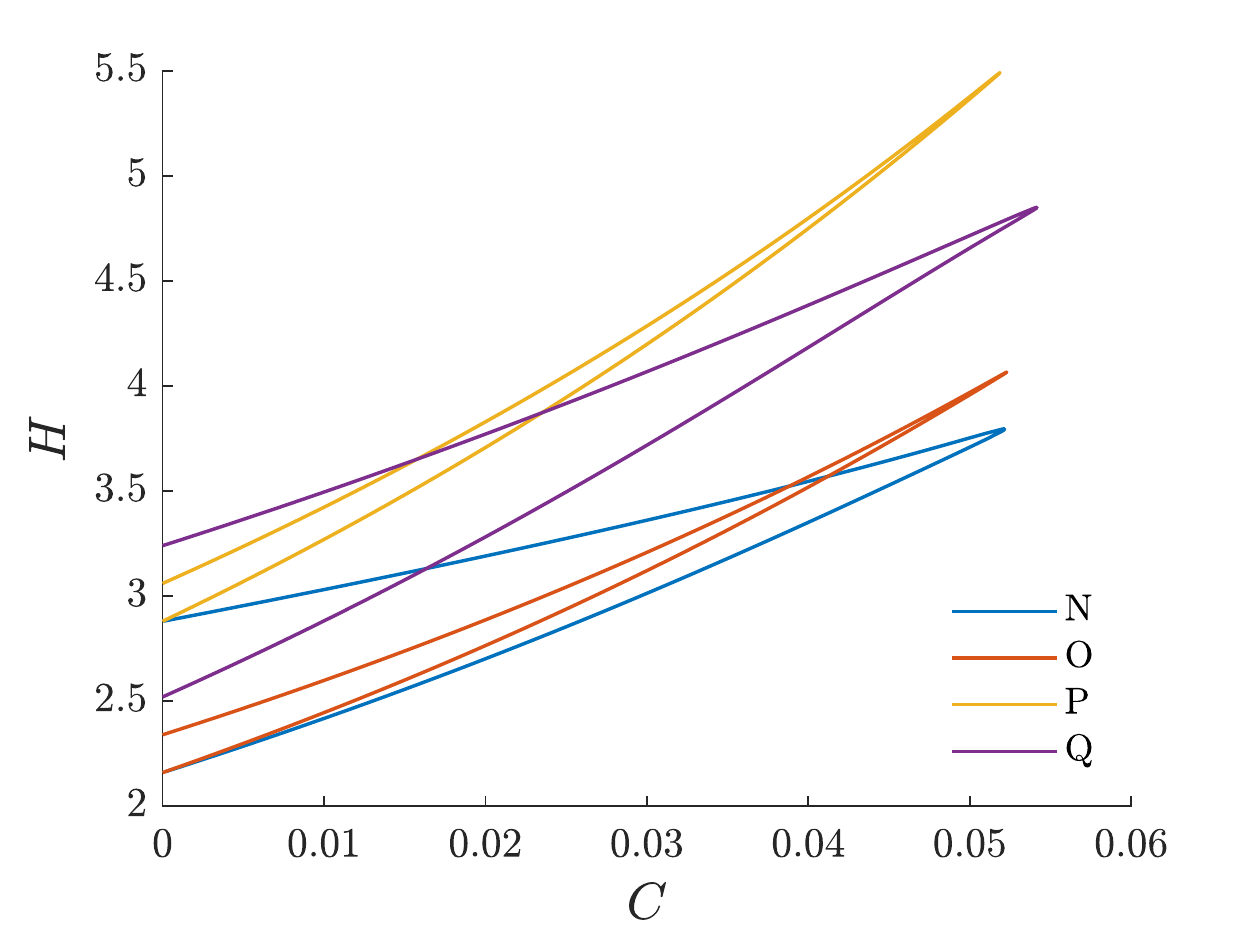} &
\includegraphics[width=.45\textwidth]{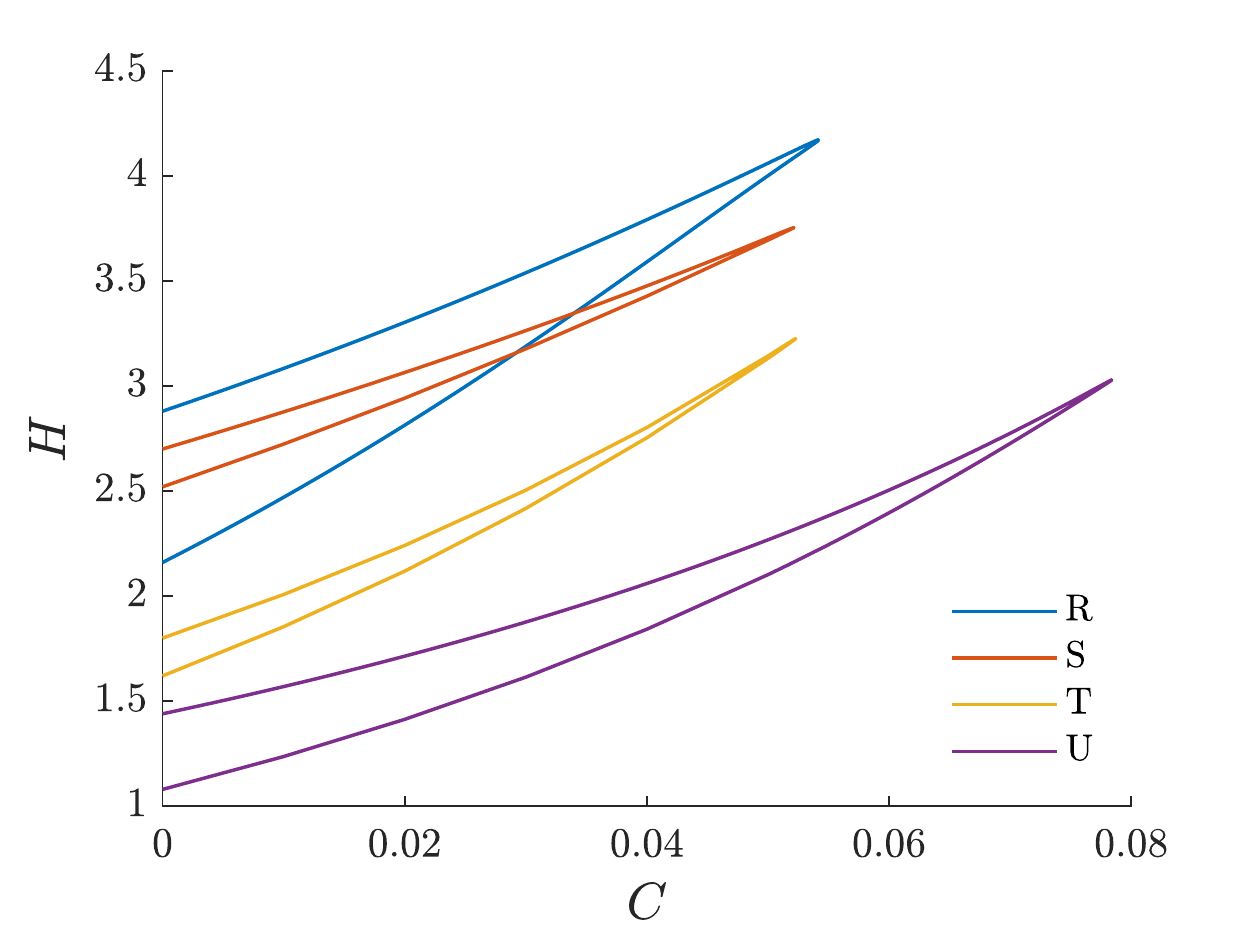} \\
\end{tabular}
\end{center}
\caption{Bifurcation diagrams for breather families obtained from a deflation perturbing the whole breather. Here, and as is shown in subsequent figures,
entirely different solution families are being probed due to the different
nature of the perturbation.}
\label{fig:diagrams2}
\end{figure}

\begin{figure}[!ht]
\begin{center}
\includegraphics[width=.9\textwidth]{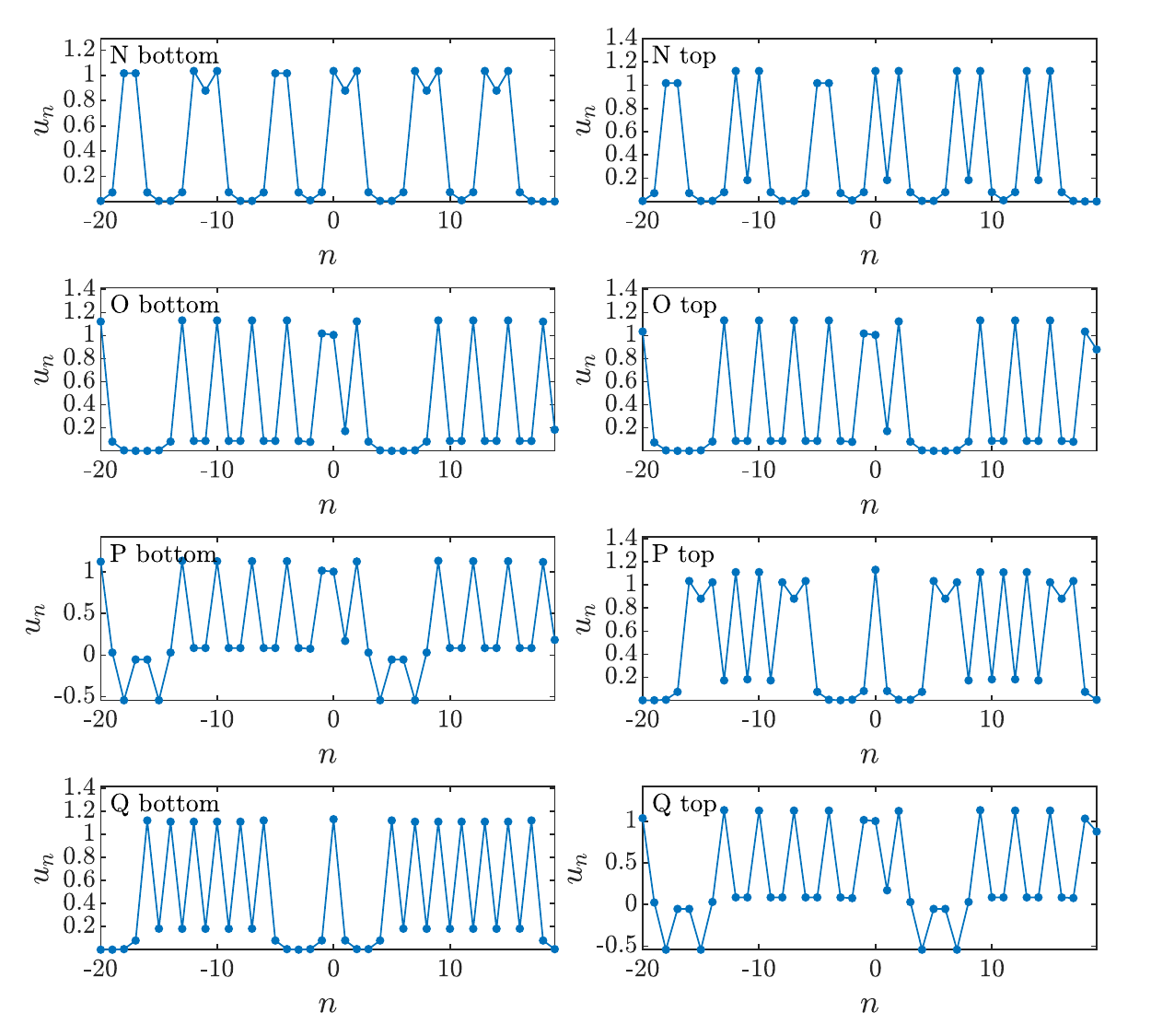}
\end{center}
\caption{Breather profiles for the solutions with $C=0.03$ in the left panel of Fig.~\ref{fig:diagrams2}. Notice the more extended nature of the
resulting solutions here and in the following figures
due to the deflation perturbation involving the
entire breather (rather than just its core).}
\label{fig:breathers2a}
\end{figure}

\begin{figure}[!ht]
\begin{center}
\includegraphics[width=.9\textwidth]{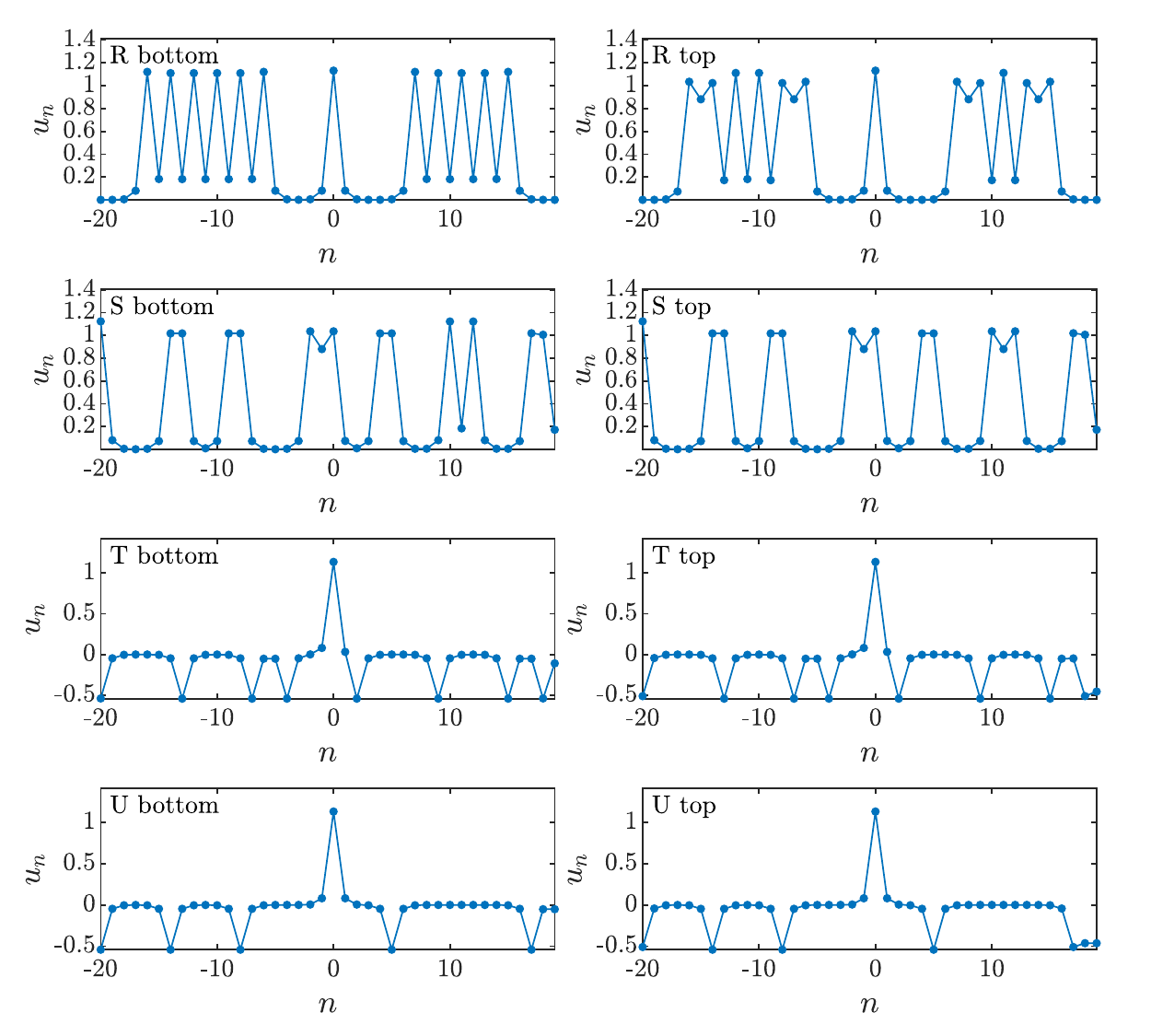}
\end{center}
\caption{Breather profiles for the solutions with $C=0.03$ in the right panel of Fig.~\ref{fig:diagrams2}.}
\label{fig:breathers2b}
\end{figure}

The approach used to find multiple solutions is outlined in Algorithm~\ref{alg:approach}. For each case listed in Tables~\ref{tab:table1} and \ref{tab:table2}, we re-initialize the list of known solutions to be the solution obtained from regular continuation, i.e.~$z^{(1)}$, and the trivial zero solution. We then perturb $z^{(1)}$ in different ways, depending on the case considered. Some cases perturb only the core of the breather (in particular, the 13 central sites), while others perturb the full breather; in the former, one mainly discovers multibreathers located at the central sites as depicted in Figs.~\ref{fig:diagrams1}-\ref{fig:breathers1d}, whereas in the latter one mainly gets multibreathers with a great number of excited sites (see Figs.~\ref{fig:diagrams2}-\ref{fig:breathers2b}).

Perturbations are made by adding a sinusoidal function to each of the relevant Fourier coefficients, i.e.,
\begin{equation}
    z_{k,n}=10^{-4}\sin(\kappa n)+z_{k,n}^{(1)}\,,
\end{equation}
with $k=0,\,1$ or $k=0,\ldots,K$.
With the perturbed initial guess at hand, we now proceed to the main deflation loop. We deflate the solutions in the list of known solutions, using the deflation operator specified in Tables ~\ref{tab:table1} and \ref{tab:table2}, and attempt to solve the deflated problem from the perturbed initial guess using the Levenberg--Marquardt algorithm. If this process does not converge, we terminate the deflation loop and move on to the next case. If this process does converge, since the Levenberg--Marquardt algorithm iterates to a minimum of the norm, a refinement with another algorithm is required in order to get a zero in the norm. To this aim we refine the new solution found by solving the original (undeflated) problem using the trust-region dogleg method. We append the refined solution to the list of known solutions and repeat, up to a maximum of ten times.

With this approach, we have found a plethora of solutions at $C=0.05$, among which we have discarded
those that correspond to the same breather solution (modulo space shift or time-reversal / space-reversal symmetry).
As indicated, aspects such as incorporation of system symmetries within
deflation~\cite{charalampidis2019bifurcation} or possibly the usage
of the energy-based approach (e.g., for space- or time-reversal) are
relevant towards addressing this aspect, a topic that we believe
will benefit further from future work.
Subsequently, we perform regular continuations by increasing and decreasing $C$
in order to get the full branch of the associated solutions, which starts at the anti-continuum limit and finishes, typically, at a turning point at $C=C_t$. In order to get the complementary branch around the turning point, we have performed continuation in the energy $H$ (\ref{eq:ham}).
To this aim, we augment the system (\ref{eq:KG_Fourier}) with
an extra equation $F_H$, given by
\begin{equation}
    F_H\equiv H(z)-H
\end{equation}
where $H$ is the energy of the solution we want to obtain, $z$ is the seed and $H(z)$ is the energy of the seed found by applying (\ref{eq:ham}); in addition, the set of variables must be augmented by including $C$, which plays the role of a Lagrange multiplier. Once a solution of the complementary branch is found, regular continuation on $C$ can be performed for this branch as well.

A summary of the bifurcation diagrams is depicted in Figures~\ref{fig:diagrams1} and \ref{fig:diagrams2}, with the former corresponding to solutions found by perturbing the breather core, and the latter to perturbations of the whole breather.
Notice that here we provide bifurcation diagrams illustrating the energy
of the obtained solutions $H$ as a function of the coupling strength
$C$; yet, it is equally possible to perform continuation
on the frequency of the solution $\omega$. The latter
is of particular interest, as well, since the slope of the respective
curves has a bearing on the breather stability (and corresponding changes
of monotonicity amount to changes of stability~\cite{DEP}).
The profile of the breather of each branch at $C=0.03$ is shown in Figs.~\ref{fig:breathers1a}-\ref{fig:breathers1d} and Figs.~\ref{fig:breathers2a}-\ref{fig:breathers2b}. Notice that, in spite of
the fact that the deflations are performed from the solutions at $C=0.05$,
we have chosen to depict them at $C=0.03$ because for the former value, in some of the branches it is hard to discern the main and the complementary branches of solutions. This is because the former value of $C$ may be quite close to the turning point value $C_t$.

It is interesting to observe the branches that appear to collide
and disappear hand-in-hand in saddle-center bifurcations in these
diagrams. For instance, the top and bottom branches of family A
can be seen to involve a pair of anti-phase site excitations to
the central-most one, with the only difference being that these excitations
occur only at the nearest neighbors to the center or also at the
next-nearest neighbors. In a similar vein, for branch B, bottom and top
both feature a core of 4 central-most sites, yet as the coupling
$C$ is increased
these move from the former towards the latter (in particular, the 2nd and the 4th
site of the bottom branch increase in amplitude to assume the amplitudes
of the top one, resulting in the collision and pairwise disappearance of
these branches). Similar patterns can be detected for all the branches that
we have considered (e.g., in branch C again an additional site to the left of
$n=0$ is excited on the bottom to make it resemble more to the top etc.).
It is worth noticing that families L and M comprise three branches, as the branch including the deflated solution bifurcates with the complementary branch through a pitchfork and the emerging branch terminates with the complementary branch through a saddle-center bifurcation.
This behaviour, which is typical of multipeaked breathers, was thoroughly described in \cite{Ross}. Let us also remark that, in every solution family, the branch encompassing the deflated solutions is the one at the bottom, except for the L and M families, where this is the one at the middle.
As discussed above, the method followed for obtaining each deflated solution is summarized in Tables~\ref{tab:table1} and \ref{tab:table2}.
Finally, it is relevant to note that the multibreathers in Figs.~\ref{fig:breathers2a}-\ref{fig:breathers2b} are rather extended
in nature, and in some cases they can be considered as a superposition of multiple single-site or two-site breather solutions. This is due to the feature that, as mentioned above, in this setting
the entire breather (rather than the core) is perturbed when performing
deflation, enabling in this way the identification of extended multibreather
solutions. Indeed, we obtain these without the expert knowledge needed
to craft specialized initial guesses that would be suitably tailored
to such waveforms. This is one of the major advantages of the ability
of this deflation method, due to its elimination property (of a previous
solution), to probe more widely, than other methods we are familiar with,
the solution landscape.

\begin{table}[!htb]
\centering
\caption{Deflation method for each solution obtained by perturbing the breather core}
\begin{tabular}{ccccc}
\hline
Family & Deflation method & Deflation order ($\nu$) & Perturbation parameter ($\kappa$) & Perturbed Fourier coefficients \\
\hline
A & Fourier & 2 & $2$ & $k=0,\,1$ \\
B & Energy & 1 & $3$ & $k=0,\,1$ \\
C & Energy & 1 & $2$ & $k=0,\ldots,K$ \\
D & Fourier & 1 & $2$ & $k=0,\,1$ \\
E & Fourier & 2 & $3$ & $k=0,\,1$ \\
F & Fourier & 1 & $1$ & $k=0,\,1$ \\
G & Energy & 1 & $2$ & $k=0,\,1$ \\
H & Energy & 1 & $1$ & $k=0,\ldots,K$ \\
I & Fourier & 4 & $3$ & $k=0,\,1$ \\
J & Energy & 2 & $1$ & $k=0,\,1$ \\
K & Fourier & 1 & $3$ & $k=0,\,1$ \\
L & Fourier & 3 & $1$ & $k=0,\,1$ \\
M & Fourier & 3 & $2$ & $k=0,\,1$ \\
\hline \\
\end{tabular}
\label{tab:table1}
\end{table}

\begin{table}[!htb]
\centering
\caption{Deflation method for each solution obtained by perturbing the whole breather}
\begin{tabular}{ccccc}
\hline
Family & Deflation method & Deflation order ($\nu$) & Perturbation parameter ($\kappa$) & Perturbed Fourier coefficients \\
\hline
N & Fourier & 2 & $1$ & $k=0,\,1$ \\
O & Fourier & 5 & $2$ & $k=0,\,1$ \\
P & Energy & 1 & $2$ & $k=0,\,1$ \\
Q & Fourier & 1 & $3$ & $k=0,\,1$ \\
R & Energy  & 1 & $3$ & $k=0,\,1$ \\
S & Fourier & 5 & $1$ & $k=0,\,1$ \\
T & Energy & 1 & $2$ & $k=0,\ldots,K$ \\
U & Energy & 1 & $1$ & $k=0,\ldots,K$ \\
\hline \\
\end{tabular}
\label{tab:table2}
\end{table}

\section{Conclusion and Future Challenges}

In the present work we have revisited the method of deflation,
which has gained considerable traction as a tool
for identifying stationary states of PDEs and for constructing their
corresponding bifurcation diagrams. Here, we proposed a way to adapt
this method towards the identification of periodic orbits in nonlinear
lattice dynamical systems with a large number of degrees of freedom.
Exploiting a Fourier decomposition renders the problem an
algebraic (nonlinear) one in the space of Fourier modes and their
spatial (lattice node) and frequency (temporal) dependence. This, in turn,
enabled us to formulate different possible variants of deflation.
We proposed two deflation operators,
deflating either the vector of Fourier coefficients,
or the spatial profile of the energy density. We argued that
the latter holds some promise towards the direction of eliminating
some of the less meaningful additional solutions
(such as the ones emerging due to
spatial parity of time reversal). We also investigated different
variants of the perturbation, upon the identification of a periodic
state, either by modifying the core region or by modifying the entire
breather. These ideas were combined successfully to
discover a wide range of outputs and a diverse array of discrete breather
families. While expert knowledge (e.g., of the anti-continuum limit)
may, in principle, be used to produce several of these families
for the Klein--Gordon lattices of interest herein, the advantage of
deflation is that one does not need such knowledge; instead, the method
can be generically applied in problems where such knowledge is not available.

Naturally, these findings merely pave the way for a considerable
array of possible further explorations in the future. On the one
hand, there are technical amendments to be considered for the method
such as the development of deflation operators that automatically eliminate
the possibility of converging to symmetry-equivalent solutions to an
existing one (lattice shifts, parity reversals, time-reversals etc.~\cite{champneys2007}).
On the other hand, it would be very interesting to assess how well
this technology works in other types of nonlinear lattices (soft- vs.~hard-nonlinearities
for example~\cite{Aubry2006,Flach2008}) and
also in higher dimensions, as well as continuous systems. Importantly,
and although we focused on a Hamiltonian system herein, it should
be highlighted that the essence of our technique can be brought to
bear independently of the conservative or non-conservative nature
of the system. Of course in the latter case, e.g.,
the energy density perspective of
deflation would not be suitable, nor would the
energy-based continuation. Nevertheless, working alternatives of
both elements exist (in the Fourier coefficient-based deflation and
the pseudo-arclength continuation, respectively), hence the extension
seems direct, although of course it needs to be suitably tested and
benchmarked. Some of these directions are currently in progress and
will be reported in future publications.

{\it Acknowledgments.}
This material is based upon work supported by the US National Science Foundation under Grants DMS-2204702 and PHY-2110030 (P.G.K.). J.C.-M.~acknowledges support from the EU (FEDER program 2014-2020) through both Consejería de Economía, Conocimiento, Empresas y Universidad de la Junta de Andalucía (under the project US-1380977), and MCIN/AEI/10.13039/501100011033 (under the projects PID2019-110430GB-C21 and PID2020-112620GB-I00). P.E.F.~acknowledges support from the UK Engineering and Physical Sciences Research Council (under grants EP/W026163/1 and EP/R029423/1). F.R.V.~acknowledges support from Consejer\'ia de Econom\'ia y Conocimiento, Junta de Andaluc\'ia (under Project RoCoSoyCo UMA18-FEDERJA-248).


\begin{thebibliography}{99}

\bibitem{willis}
S.~Flach and C.R. Willis.
\newblock {Physics Reports} {\bf 295}, 181 (1998).

\bibitem{Flach2008}
S. Flach and A.V. Gorbach.
\newblock {Physics Reports} {\bf 467}, 1 (2008).

\bibitem{Aubry2006}
S.~Aubry.
\newblock {Physica D} {\bf 216}, 1 (2006).

\bibitem{takeno}
A.~J. Sievers and S.~Takeno.
\newblock {Phys. Rev. Lett.} {\bf 61}, 970 (1988).

\bibitem{page}
J.~B. Page.
\newblock {Phys. Rev. B} {\bf 41}, 7835 (1990).

\bibitem{mackay}
R~S MacKay and S~Aubry.
\newblock {Nonlinearity}, {\bf 7}, 1623 (1994).

\bibitem{LEDERER20081}
F. Lederer, G.I. Stegeman, D.N. Christodoulides, G. Assanto,  M. Segev, and Y. Silberberg.
\newblock {Physics Reports} {\bf 463}, 1 (2008).

\bibitem{RevModPhys.78.179}
O. Morsch and M. Oberthaler.
\newblock {Rev. Mod. Phys.} {\bf 78}, 179 (2006).

\bibitem{Chong2018}
C. Chong and P.G. Kevrekidis.
\newblock {\em {Coherent Structures in Granular Crystals: From Experiment and
  Modelling to Computation and Mathematical Analysis}}.
\newblock Springer International Publishing, 2018.

\bibitem{remoissenet}
M.~Remoissenet.
\newblock {\em Waves Called Solitons}.
\newblock Springer-Verlag, Berlin, 1999.

\bibitem{alex}
P.~Binder, D.~Abraimov, A.~V. Ustinov, S.~Flach, and Y.~Zolotaryuk.
\newblock {Phys. Rev. Lett.} {\bf 84}, 745 (2000).

\bibitem{alex2}
E.~Tr\'{\i}as, J.~J. Mazo, and T.~P. Orlando.
\newblock {Phys. Rev. Lett.} {\bf 84}, 741 (2000).

\bibitem{cantilevers}
M. Sato, B.E. Hubbard, A.J. Sievers. Rev. Mod. Phys. {\bf 78}, 137 (2006)

\bibitem{NbNi}
M. Haas, V. Hizhnyakov, A. Shelkan, M. Klopov, A.J. Sievers, Phys. Rev. B {\bf 84}, 144303
(2011)

\bibitem{carbon}
J.A. Baimova, E.A. Korznikova, I.P. Lobzenko, S.V. Dmitriev, Rev. Adv. Mater. Sci. {\bf 42}, 68
(2015)

\bibitem{Peybi}
M.~Peyrard.
\newblock {Nonlinearity} {\bf 17}, R1 (2004).

\bibitem{farrell2015deflation}
P.E. Farrell, A. Birkisson, and S.W. Funke.
\newblock {SIAM J. Sci. Comput.} {\bf 37}, A2026 (2015).

\bibitem{charalampidis2018computing}
E.G. Charalampidis, P.G. Kevrekidis, and P.E. Farrell.
\newblock {Commun. Nonlinear Sci. Numer. Simulat.} {\bf 54}, 482 (2018).

\bibitem{charalampidis2019bifurcation}
E.G. Charalampidis, N. Boull{\'e}, P.E. Farrell, and P.G. Kevrekidis.
\newblock {Commun. Nonlinear Sci. Numer. Simulat.} {\bf 87}, 105255 (2020).

\bibitem{boulle2020deflation}
N. Boull{\'e}, E.G. Charalampidis, P.E. Farrell, and P.G. Kevrekidis.
\newblock {Phys. Rev. A} {\bf 102}, 053307 (2020).

\bibitem{stereo}
N.~Boull\'e, I.~Newell, P.~E. Farrell, and P.~G. Kevrekidis.
\newblock {Phys. Rev. A} {\bf 107}, 012813 (2023).

\bibitem{vraha}
V.~S. Kalantonis, E.~A. Perdios, A.~E. Perdiou, O.~Ragos, and M.~N. Vrahatis.
\newblock {Astrophysics and Space Science}, {\bf 288}, 489 (2003).

\bibitem{Marin}
J.L. Marín and S.~Aubry.
\newblock Nonlinearity {\bf 9}, 1501 (1996).

\bibitem{Aubry97}
S. Aubry, Physica D {\bf 103}, 201 (1997).

\bibitem{phantom}
A.M. Morgante, M. Johansson, S. Aubry, G. Kopidakis, J. Phys. A: Math. Gen. {\bf 35}, 4999
(2002).

\bibitem{Juan}
J.F.R. Archilla, J.~Cuevas, B.~Sánchez-Rey, and A.~Alvarez.
\newblock {Physica D} {\bf 180}, 235 (2003).

\bibitem{Koukouloyannis_2009}
V. Koukouloyannis and P.G. Kevrekidis.
\newblock {Nonlinearity} {\bf 22}, 2269 (2009).

\bibitem{Sakovich}
D.E. Pelinovsky, A. Sakovich, Nonlinearity {\bf 25}, 3423 (2012).

\bibitem{DEP}
P.G. Kevrekidis, J. Cuevas-Maraver, and D.E. Pelinovsky.
\newblock {Phys. Rev. Lett.} {\bf 117}, 094101 (2016).

\bibitem{Ross}
R. Parker, J. Cuevas-Maraver, P.G. Kevrekidis, A. Aceves.
\newblock {Nonlinearity} {\bf 35}, 5714 (2022).

\bibitem{champneys2007}
A. R. Champneys, B. Sandstede.
\newblock {In \emph{Numerical Continuation Methods for Dynamical Systems}}.
\newblock Springer Dordrecht, 2007.

\end{thebibliography}
\end{document}